\documentstyle[12pt]{article}
\begin{document}
\title{\bf Drag Force of Moving Quark at $\mathcal{N}$=2 Supergravity}
\author{J. Sadeghi $^{a,}$\thanks{Email: pouriya@ipm.ir}\hspace{1mm}
and B.Pourhassan$^{a}$\thanks{Email: b.pourhassan@umz.ac.ir}\\
$^a$ {\small {\em  Sciences Faculty, Department of Physics, Mazandaran University,}}\\
{\small {\em P .O .Box 47415-416, Babolsar, Iran}}}
\maketitle
\begin{abstract}
\noindent In this paper we consider a moving quark in the thermal
plasma at the $\mathcal{N}$=2 Supergravity theory. By using the
$AdS$/CFT correspondence we obtain energy loss of the quark. Then we
consider the higher derivative corrections in charged $AdS$-black
hole and calculate the drag force on the moving quark in the thermal
plasma. Also we find a limit which $\mathcal{N}$=2 Supergravity
solutions are corresponding to the $\mathcal{N}$=4 Super Yang-Mills
solutions for the heavy quark.
\end{abstract}
\section{Introduction}
The relation between gauge theories and string theory has been the
subject of many important researches  in the last three decades.
First, Maldacena in his paper [1] proposed $AdS$/CFT correspondence,
therefore the $AdS$/CFT correspondence sometimes called Maldacena
duality. According to this conjecture there is a relation between a
conformal field theory (CFT) in $d$-dimension and a supergravity
theory in $d+1$-dimensional anti de Sitter ($AdS$) space. Maldacena
suggests that a quantum string in $d+1$-dimensional $AdS$ space,
mathematically is equivalent to the ordinary quantum field theory
with conformal invariance in $d$-dimensional space-time which lives
on the boundary of $AdS_{d+1}$ space. Some details in preliminary
formulation of Maldacena are explained by independent works of
Witten [2] and Gubser,  et al. [3]. An example of $AdS$/CFT
correspondence is the relation between type IIB string theory in
$AdS_{5}\times S^{5}$ space and $\mathcal{N}$=4 super Yang-Mills
gauge theory on the 4-dimensional boundary of $AdS_{5}$ space.
Today, one of the important issues of researches is using $AdS$/CFT
correspondence in many complicated problem of QCD. For example most
of the  research is about energy loss of moving charged particles in
plasma based on week coupling [4-15]. But if one would like to
understand the dynamics of such systems in the strong coupling, then
there are some complicated calculations in QCD. Therefore energy
loss of moving quark through the $\mathcal{N}$=4 super Yang-Mills
thermal plasma [16] is studied by using $AdS$/CFT correspondence
[17-20]. Furthermore the drag force on a pair of quark-anti quark is
considered [17, 21, 22, 23]. In this way adding the temperature to
the system in the gauge theory is corresponding to introduce  a
black hole (black brane) in the center of  $AdS_{5}$ space. In this
model, at non-zero temperature, one can image open string stretched
from D-brane to the horizon and end point of string on D-brane
represents a quark, so the quark moves and pulls the string. By
study the behavior of the string end point, we can obtain energy
loss of the
quark and drag force in the gauge theory .\\
In this paper we consider the moving quark in $\mathcal{N}$=2
supergravity thermal medium [24, 25], and calculate drag force  in
various situations. Indeed, solutions of $\mathcal{N}$=2
supergravity may be solutions of supergravity  theory with more
supersymmetry ($\mathcal{N}$=4 and $\mathcal{N}$=8). The
$\mathcal{N}$=2 supergravity theory in five dimensions can be
obtained by compactifying eleven dimensional supergravity in a
3-fold Calabi-Yau [26]. Also anti de Sitter supergravity can obtain
by gauging the $U(1)$ subgroup of the $SU(2)$ group in
$\mathcal{N}$=2 supersymmetric algebra. Also we would like to add a
constant $B$ field to the system and find effect of constant
electric and magnetic field on the drag force.
Already the drag force in a thermal plasma of $\mathcal{N}$=4 super Yang-Mills theory under the influence of
non-zero NSNS $B$-field background has been studied [27].\\
Then we consider higher derivative corrections to $AdS_{5}$ charged
black hole and obtain drag force. Already the higher-derivative
curvature corrections to type IIB supergravity was done [28-31].
Also effect of curvature-squared corrections on the drag force of
moving heavy quark in the $\mathcal{N}$=4 super Yang-Mills plasma is
considered by Ref.[32]. Furthermore presence of $R^{2}$-term in
curvature tensor in $\mathcal{N}$=2 supergravity theory has been
studied [33]. However, we use the solutions of spherical symmetric
$AdS_{5}$ charged black hole [24, 25] and find drag force on the
moving quark through thermal plasma and then consider the effect of
higher derivative terms [34] on the drag force. We note that the
stated analysis might be invalid in lower dimension with
higher-derivative effective action, without considering proper
Kaluza-Klein (K-K) reduction. Therefore universality can be
established by carefully analyzing Kaluza-Klein
reduction of ten-dimensional action [35].\\
This paper is organized as follows, in  section 2 we review a model
for the drag force on quark. Then in section 3, by using solutions
of charged $AdS$-black hole, we find string equation of motion and
solve it in three interesting case.  In section 4 we discuss the
quasi normal modes of string without any external field. In section
5 we add B-field to the system and discuss about the effect of
constant electric and magnetic fields, as an external fields, on the
drag force. In such cases we follow similar methods used in [17, 20,
27] directly. Finally in section 6 we consider effect of higher
derivative terms in our solutions and give some results and
summaries in section 7.
\section{ Drag Force }
We consider a particle which moves in thermal medium with viscosity,
therefore it senses a drag force due to medium. If we consider $p$
as momentum of the particle which moves under external force $F$ and
friction coefficient $\mu$, one can write equation of motion as
[17],
\begin{equation}\label{s1}
\dot{p}=F-\mu p.
\end{equation}
In order to obtain some information about drag force, it is useful
to consider two special cases. First, we assume that momentum of the
particle is constant, hence we have $F=\mu p$ and for particle with
mass $m$ and non-relativistic momentum $p=mv$ one obtain $\mu
m=\frac{F}{v}$. So, by measurement of velocity of particle for given
force we can obtain $\mu m$. It shows that we can't find $\mu$ independently. The parameter $\mu m$ is called drag force coefficient.\\
In the second case we assume that external force does not exist, so
from equation of motion (1) one can find $p(t)=p(0)e^{-\mu t}$. In
other words by measurement of ratio $\frac{\dot{p}}{p}$ or
$\frac{\dot{v}}{v}$ we can determine $\mu$ without any dependance to
$m$. These lead us to obtain drag force for a moving quark in plasma [17].\\
The starting point in calculating drag force on quark, is the string
action,
\begin{equation}\label{s2}
S=T_{0}\int{d^{2}\sigma\mathcal{L}}.
\end{equation}
By using the Euler-Lagrange differential equation
$\frac{\partial}{\partial\xi}-\frac{d}{dt}\frac{\partial\mathcal{L}}{\partial\xi^{\prime}}=0$
with respect to string coordinate $\xi$ and canonical momentum
density
$\pi_{\xi}=\frac{\partial\mathcal{L}}{\partial\xi^{\prime}}$, we
have the following equation,
\begin{equation}\label{s3}
\frac{dp}{dt}=\frac{1}{v}\frac{dE}{dt}=\frac{dE}{dx}=-T_{0}{\pi}_{\xi},
\end{equation}
this shows that the rate of energy and momentum loss is proportional
to the canonical momentum density. The quark with Lagrangian mass
$m$ floats in the thermal plasma with temperature $T$, and it has a
rest mass $M_{rest}(T)$ which is different from physical
(Lagrangian) mass of the quark, because the temperature effect on
the mass. On the other hand if the quark moves in the thermal plasma
then it will have a kinetic mass $M_{kin}(T)$, so we have [17],
\begin{equation}\label{s4}
E=M_{rest}(T)+\frac{p^{2}}{2M_{kin}(T)}.
\end{equation}
Difference of $M_{kin}(T)$ and $M_{rest}(T)$ for heavy quark
($m>>\Delta m(T)$) is negligible, so we have,
\begin{equation}\label{s5}
M_{kin}(T)=M_{rest}(T)+{\mathcal{O}}\left[m(\frac{\triangle
m(T)}{m})^{2}\right],
\end{equation}
where $\Delta m(T)$ is thermal mass shift. By knowledge about the
drag force and mass we can obtain diffusion coefficient for
non-relativistic quark, which depends to temperature and drag force
[17],
\begin{equation}\label{s6}
D=\frac{T}{m\mu},
\end{equation}
where parameter $D$ is a phenomenological property.
\section{ String Equation of Motion}
As we know the temperature in the supersymmetric gauge theory is
equal to existence a black hole with a flat horizon in the anti de
Sitter space. As mentioned [33] Lagrangian for the bosonic sector of
pure $\mathcal{N}$=2 gauged supergravity in five dimensions is,
\begin{equation}\label{s7}
e^{-1}{\mathcal{L}}_{0}=R-\frac{1}{4}F_{\mu\nu}F^{\mu\nu}+12\Lambda^{2}+\frac{1}{12\sqrt{3}}\epsilon^{\mu\nu\rho\sigma\lambda}F_{\mu\nu}F_{\rho\sigma}A_{\lambda}.
\end{equation}
Then the $AdS_{5}$
 black hole solutions, given by [24, 25],
\begin{eqnarray}\label{s8}
ds^{2}&=&-\frac{f}{H^{2}}dt^{2}+H(r^{2}dx^{2}+\frac{dr^{2}}{f}),\nonumber\\
A&=&\sqrt{3}\coth\beta(\frac{1}{H}-1)dt,\nonumber\\
f&=&1-\frac{\eta}{r^{2}}+\Lambda^{2}r^{2}H^{3} ,\nonumber\\
H&=&1+\frac{\eta \sinh^{2}\beta}{r^{2}},
\end{eqnarray}
where $r$ is axis along the black hole, so the horizon of black hole
is in $r=r_{h}$. Also we assume that the motion in sphere $S^{5}$ is
only in transverse axis $x$. The $\beta$ parameter is related to the
electric charge of black hole and $\Lambda$ is cosmological
constant, also expansion parameter $\eta$ is called
non-extremality parameter.\\
The dynamics of a open string is described by the following
Nambo-Goto action,
\begin{equation}\label{s9}
S=-T_{0}\int{d\tau d\sigma \sqrt{-g}},
\end{equation}
where $-g=-det{g_{ab}}$, $g_{ab}$ is the metric of string
worldsheet, and space-time metric $G_{\mu\nu}$ is given by line
element (8). As we told the string moves only in $x$ direction and
we use static gauge so that we take $\tau=t$ and $\sigma=r$.
Therefore string world-sheet described by $x(t, r)$, so we can
write,
\begin{equation}\label{s10}
-g=\frac{1}{H}-\frac{H^{2} r^{2}}{f}{\dot{x}}^{2}+\frac{f
r^{2}}{H}{x^{\prime}}^{2},
\end{equation}
and the equation of motion is,
\begin{equation}\label{s11}
\frac{\partial}{\partial r}(\frac{f
r^{2}}{H\sqrt{-g}}x^{\prime})-\frac{H^{2}r^{2}}{f}\frac{\partial}{\partial
t}(\frac{\dot{x}}{\sqrt{-g}})=0.
\end{equation}
Here, we can obtain component of canonical momentum density for
$\mu=x, r, t$ as,
\begin{eqnarray}\label{s12}
{\pi}_{t}^{0} &=& -\frac{T_{0}}{\sqrt{-g}}\frac{(1+fr^{2}{x^{\prime}}^{2})}{H},\nonumber\\
{\pi}_{x}^{0} &=& \frac{T_{0}}{\sqrt{-g}}\frac{H^{2}r^{2}}{f}\dot{x},\nonumber\\
{\pi}_{r}^{0} &=&
-\frac{T_{0}}{\sqrt{-g}}\frac{H^{2}r^{2}}{f}\dot{x}x^{\prime},
\end{eqnarray}
and
\begin{eqnarray}\label{s13}
{\pi}_{t}^{1} &=& \frac{T_{0}}{\sqrt{-g}}\frac{fr^{2}}{H}\dot{x}x^{\prime},\nonumber\\
{\pi}_{x}^{1} &=& -\frac{T_{0}}{\sqrt{-g}}\frac{fr^{2}}{H}x^{\prime},\nonumber\\
{\pi}_{r}^{1} &=&
\frac{T_{0}}{\sqrt{-g}}(-\frac{1}{H}+\frac{H^{2}r^{2}}{f}{\dot{x}}^{2}).
\end{eqnarray}
Then the total energy and momentum of string can be obtained by
following relations,
\begin{eqnarray}\label{s14}
E&=&-\int{dr{\pi}_{t}^{0}},\nonumber\\
p&=&\int{dr{\pi}_{x}^{0}}.
\end{eqnarray}
We note that there is interesting limit where above relations
(10-13) are corresponding to the case of a heavy quark moving
through $\mathcal{N}$=4 supersymmetric Yang-Mills plasma. That is
$\eta\rightarrow0$ limit where we have $H=1$, then by rescaling
$r=L^{2}u$ ($L$ is $AdS$ radius) and setting $f(r)\equiv h(u)$ (in
4-dimension we know $h=u^{2}[1-(\frac{u_{h}}{u})^{4}]$, where
$u_{h}$ is position of black hole horizon) our results reduce to the
Ref. [17].\\
The Hawking temperature of the solutions (8) is given by [36],
\begin{equation}\label{s15}
T=\frac{2+3k-k^{3}}{2(1+k)^{\frac{3}{2}}}\frac{r_{h}}{\pi L^{2}},
\end{equation}
where $k\equiv\frac{\eta \sinh^{2}\beta}{r_{h}^{2}}$. Clearly at the
$\eta\rightarrow0$ limit we have $T=\frac{r_{h}}{\pi L^{2}}$, which
is agree with Hawking temperature of the black hole for $d=4$ [17].
As we know the consideration of a quark in the gauge theory is
corresponding to string in $AdS$ space, also spiking about
temperature of plasma is corresponding to existence of a black hole
in $AdS$ space.  Similarly to this, the consideration quark flavor
in the field theory is corresponding to adding a $D$-brane.
$D$-brane covers an sphere $S^{3}$ in
$S^{5}$ which has minimum radius $r_{m}$.\\
Now we come back to string equation of motion. In order to solve the
equation of motion (11) and obtain drag force, we consider three
cases. First of all, we take an static string which is corresponding
to the rest quark. In the second case we assume that the straight
string moves with the constant velocity. Finally in the third case,
for physical motion, we
consider curved moving string with constant speed $v$.\\
A quark in the ${\mathcal{N}}$=2 supergravity thermal plasma is
equal to the string which stretched from $D$-brane to the black hole
horizon. End point of string on $D$-brane represents the quark.\\
In this case the simplest solution which satisfy equation of motion
(11) is $x(r, t)=x_{0}$. It describes an static string stretched
straightforwardly from $r=r_{m}$ on $D$-brane to $r=r_{h}$ in the
black hole horizon, and represents the rest quark in thermal plasma
clearly. Therefore by using equations (12) and (14), we can
calculate total energy and momentum.  Momentum of the rest quark
vanishes,$\pi_{x}^{0}=\pi_{r}^{0}=\pi_{t}^{1}=\pi_{x}^{1}=0$, as we
expected and the total energy will be as,
\begin{equation}\label{s16}
E=T_{0}\left(r_{m}\sqrt{1+\frac{\eta\sinh^{2}\beta}{r_{m}^{2}}}-r_{h}\sqrt{1+\frac{\eta\sinh^{2}\beta}{r_{h}^{2}}}\right).
\end{equation}
As we see in equation (15) the temperature of black hole is zero at
$ r_{h}^{2}=\frac{\eta\sinh^{2}\beta}{2}$. There are also another
solutions such as $r_{h}^{2}=0$ and $r_{h}^{2}=-\eta\sinh^{2}\beta$
which have singularity, therefore they are not allowed. Again we
note that at the limit of $\eta\rightarrow0$ the temperature is
proportional to $r_{h}$ and the zero temperature limit is equal to
$r_{h}\rightarrow0$ [17]. However we know that the extremal limit of
background (8) is obtained in $\eta\rightarrow\infty$ and
$\beta\rightarrow0$ limit [34]. On the other hand in zero
temperature  we can interpret $E$ as physical (Lagrangian) mass of
quark. Hence, in limit of zero temperature we can write,
\begin{equation}\label{s17}
E=m=T_{0}\left(r_{m}\sqrt{1+\frac{\eta\sinh^{2}\beta}{r_{m}^{2}}}-\sqrt{\frac{3\eta\sinh^{2}\beta}{2}}\right).
\end{equation}
By increasing the radius, the mass of quark increases. So that if
$D$-brane moves to boundary of $AdS$ space ($r\rightarrow \infty$)
the mass of quark will be infinite. But the increasing of the
temperature effect on the relation between Lagrangian mass and
$r_{m}$. The energy $E$ in (16) at non-zero temperature interpreted
as free energy of the rest quark in $\mathcal{N}$=2 supergravity
thermal plasma, then one can obtain the relation between free energy
and the thermal rest mass of quark. So we follow from used
techniques in [17] and summarize some results in table 1.\\
Now we assume that the straight string moves with the speed of $v$,
so $x(r, t)=x_{0}+vt$  may be solution of equation (11). In this
case one can find non-zero components of the canonical momentum
density as,
\begin{eqnarray}\label{s18}
{\pi}_{t}^{0} &=& -\frac{T_{0}}{H\sqrt{\frac{1}{H}-\frac{H^{2} r^{2}}{f}v^{2}}},\nonumber\\
{\pi}_{x}^{0} &=& \frac{T_{0}H^{2}r^{2}v}{f\sqrt{\frac{1}{H}-\frac{H^{2} r^{2}}{f}v^{2}}},\nonumber\\
{\pi}_{r}^{1} &=& -T_{0}{\sqrt{\frac{1}{H}-\frac{H^{2}
r^{2}}{f}v^{2}}}.
\end{eqnarray}
We see that the energy and $x$-component of momenum densities on the
string worldsheet are non-zero, so one can obtain total energy and
momentum of the string, but we don't like to perform this step for
following reason. As we saw, $\eta\rightarrow0$ limit of the
$\mathcal{N}$=2 supergravity theory is corresponding to the
$\mathcal{N}$=4 super yang-mills theory in the problem of drag
force, so the moving straight string is not a physical motion,
because in that case the square root quantity in the action is
negative and we have imaginary action, energy and momentum.
Therefore, we leave this state and take curved moving string with
speed of $v$ as physical motion, for such system, one can choose,
\begin{equation}\label{s19}
x(r,t) =x(r)+vt,
\end{equation}
as solution of string equation (11). First we try to determine
$x(r)$. It is clear that $\dot{x}$, $x^{\prime}$ and $\sqrt{-g}$ are
independent of time, therefore from equation of motion (11) we have,
\begin{equation}\label{s20}
x^{\prime}=\frac{Hv\sqrt{-g}C}{f r^{2}},
\end{equation}
where $C$ is a constant and,
\begin{equation}\label{s21}
-g=\frac{1}{H}-\frac{H^{2} r^{2}}{f}v^{2}+\frac{f
r^{2}}{H}{x^{\prime}}^{2}.
\end{equation}
Here, we use the equations (13) and (20), and obtain momentum
density as,
\begin{eqnarray}\label{s22}
{\pi}_{t}^{1} &=& T_{0}Cv^{2},\nonumber\\
{\pi}_{x}^{1} &=& -T_{0}Cv.
\end{eqnarray}
As we see this two  currents are constant along the string. Now by
using the equation (20) and (21) we have,
\begin{equation}\label{s23}
-g=\frac{r^{2}}{H}\frac{f-H^{3}r^{2}v^{2}}{fr^{2}-{C^{2}v^{2}H}}.
\end{equation}
The important problem here is that $-g$ must be positive. By
appropriate choice of constant $C$, we have $-g>0$ everywhere. So
one can obtain,
\begin{equation}\label{s24}
C=\pm(1+\frac{\eta\sinh^{2}\beta}{r_{c}^{2}})r_{c}^{2},
\end{equation}
where critical radius $r_{c}$ is root of following equation,
\begin{equation}\label{s25}
r^{2}-\eta+(\Lambda^{2}-v^{2})r^{4}(1+\frac{\eta\sinh^{2}\beta}{r^{2}})^{3}=0.
\end{equation}
Furthermore, by using $C$ from equation (24) in equation (22) the
energy current into the horizon is,
\begin{equation}\label{s26}
{\pi}_{t}^{1} =
T_{0}v^{2}(1+\frac{\eta\sinh^{2}\beta}{r_{c}^{2}})r_{c}^{2},
\end{equation}
and the momentum current into the horizon is,
\begin{equation}\label{s27}
{\pi}_{x}^{1} =
-T_{0}v(1+\frac{\eta\sinh^{2}\beta}{r_{c}^{2}})r_{c}^{2}.
\end{equation}
Then from equation (12) we obtain the ${\pi}_{t}^{0}$ and
${\pi}_{x}^{0}$ as follow,
\begin{eqnarray}\label{s28}
{\pi}_{t}^{0} &=& -\frac{T_{0}}{(1+\frac{\eta\sinh^{2}\beta}{r^{2}})}\sqrt{\frac{(1+\frac{\eta\sinh^{2}\beta}{r_{c}^{2}})^{2}r_{c}^{4}}{(1+\frac{\eta\sinh^{2}\beta}{r_{h}^{2}})r_{h}^{4}}}\left[1+\frac{\left((1+\frac{\eta\sinh^{2}\beta}{r^{2}})v\sqrt{(1+\frac{\eta\sinh^{2}\beta}{r_{h}^{2}})}r_{h}^{4}\right)^{2}}{\left(1-\frac{\eta}{r^{2}}+\Lambda^{2}r^{2}(1+\frac{\eta\sinh^{2}\beta}{r^{2}})^{3}\right)r}\right]\nonumber\\
{\pi}_{x}^{0} &=&
T_{0}\sqrt{\frac{(1+\frac{\eta\sinh^{2}\beta}{r_{c}^{2}})^{2}r_{c}^{4}}{(1+\frac{\eta\sinh^{2}\beta}{r_{h}^{2}})r_{h}^{4}}}
\frac{(1+\frac{\eta\sinh^{2}\beta}{r^{2}})^{2}r^{2}v}{1-\frac{\eta}{r^{2}}+\Lambda^{2}r^{2}(1+\frac{\eta\sinh^{2}\beta}{r^{2}})^{3}},
\end{eqnarray}
in order to obtain the total energy and momentum of string one can
integrate them over the worldsheet from $r_{h}$ to $r_{m}$. We note
that for $\eta\rightarrow0$ limit, the critical radius is
$r_{c}^{2}=\frac{1}{v^{2}-\Lambda^{2}}$, therefore equation (24)
reduces to,
\begin{equation}\label{s29}
C=\pm\frac{1}{v^{2}-\Lambda^{2}},
\end{equation}
and we have,
\begin{equation}\label{s30}
-g=r_{h}^{4}(v^{2}-\Lambda^{2})^{2}.
\end{equation}
We saw that loss of the energy current of quark by the string,
$\pi_{t}^{1}$, is proportional to $C$ ( see equation (22)). So the
positive sign in $C$ shows the energy current from quark to the
horizon along string. But the negative sign of $C$ shows energy
current from the horizon to quark, in this case the string moves and
pull the quark, so it is non-physical situation, therefore we give
positive sign in equations (24) and (29) for constant $C$. Hence by
using equations (20), (29) and (30) we obtain $x^{\prime}$ as,
\begin{equation}\label{s31}
x^{\prime}=\frac{vr_{h}^{2}}{(1+\Lambda^{2}r^{2})r^{2}}.
\end{equation}
Here we should strongly recommend that at $\eta\rightarrow0$ limit
one can compare presence problem to the quark moving through
${\mathcal{N}}=4$ SYM plasma. As mentioned already we have
correspondence between ${\mathcal{N}}=2$ supergravity and
${\mathcal{N}}=4$ super Yang-Mills Theories. Hence by integrating
from $x^{\prime}$ with respect to $r$, we find string coordinate
(19) as,
\begin{equation}\label{s32}
x(r,t)
=x_{0}+vt+vr_{h}^{2}\left[\frac{\pi}{2}-\Lambda\tan^{-1}(\Lambda
r)-\frac{1}{r}\right],
\end{equation}
which, other than a constant coefficient, is different from [17]
only in the last term. It is due to difference between $f(r)$ in our
paper and $h(u)$ in ${\mathcal{N}}=4$ SYM. The energy and momentum
current to the black hole horizon from equation (22) give us the
following equations,
\begin{eqnarray}\label{s33}
{\pi}_{t}^{1} &=& \frac{T_{0}v^{2}}{v^{2}-\Lambda^{2}},\nonumber\\
{\pi}_{x}^{1} &=& -\frac{T_{0}v}{v^{2}-\Lambda^{2}}.
\end{eqnarray}
Now we would like to interpret solutions of stationary string as
describing the steady-state behavior of a moving quark through the
${\mathcal{N}}=2$ supergravity medium under effect of a constant
electric field $\varepsilon$. The velocity of quark is going to an
equilibrium value $v$ where the drag force on quark is equal to the
external force due to the constant electric field. The constant
electric field $\varepsilon$ is a real parameter. That is a $U(1)$
gauge which coupled to the brane. The work of electric field on the
quark is proportional to its velocity i.e. $\varepsilon\cdot v$.
This work is equal to the energy loss of quark in the plasma.
According to the relation (33)  the energy current of the string
through the horizon is,
\begin{equation}\label{s34}
{\pi}_{t}^{1} =-{\pi}_{x}^{1}v =
T_{0}v^{2}(1+\frac{\eta\sinh^{2}\beta}{r_{c}^{2}})r_{c}^{2},
\end{equation}
where in  $\eta\rightarrow0$ limit reduces to the
${\pi}_{t}^{1}=\frac{T_{0}v^{2}}{v^{2}-\Lambda^{2}}$. Therefore, one
can write equation ${\pi}_{x}^{1}=-\varepsilon$ and find,
\begin{equation}\label{s35}
\mu m=T_{0}(1+\frac{\eta\sinh^{2}\beta}{r_{c}^{2}})r_{c}^{2},
\end{equation}
where we assume that the quark is an excitation mode of string with
mass $m$ and non-relativistic momentum $p=mv$, so we can gave the
rate of transferred momentum ${\pi}_{x}^{1}$ equal to momentum loss
of quark $\dot{p}=-\mu p$. Therefore, at the  $\eta\rightarrow0$
limit for heavy quark (low velocity) one can obtain,
$\dot{p}=\frac{T_{0}}{\Lambda^{2}}$. Indeed,
$\frac{dp}{dt}=-{\pi}_{x}^{1}$ and $\frac{dE}{dt}={\pi}_{t}^{1}$ are
the rate of momentum and energy, respectively, which transfer to the
quark by the electric field. By using non-relativistic momentum
$p=mv$, one can write the diffusion coefficient for the quark as,
\begin{equation}\label{s36}
D=\frac{T}{T_{0}}\left[(1+\frac{\eta\sinh^{2}\beta}{r_{c}^{2}})r_{c}^{2}\right]^{-1}.
\end{equation}
As we see, information about the drag force is equivalent to
information about the diffusion coefficient of quark. One can take
$\eta\rightarrow0$ limit in the above expression and neglect the
squared-velocity terms for heavy quark and obtain
$D=-\frac{T}{T_{0}}\Lambda^{2}.$
\section{Small Fluctuations}
In this section we again consider  a  quark moving in the
$\mathcal{N}$=2 supergravity thermal plasma without any external
field. The aim of this section is studying behavior of string after
long time and with slow velocity. We consider the dynamics of such a
system at a long time as an small perturbation in the static string
which describes the rest of quark. Therefore, we have quasinormal
modes of string worldsheet which are small fluctuations in the
string. This small fluctuations around the string means that
$\dot{x}$ and $x^{\prime}$ in string equation of motion are small.
Hence, we neglect $\dot{x}^{2}$ and ${x^{\prime}}^{2}$ in $-g$, this
leads us to  $-g=\frac{1}{H}$, then equation of motion (11) reduces
to,
\begin{equation}\label{s37}
\frac{\partial}{\partial r}(\frac{f
r^{2}}{\sqrt{H}}x^{\prime})-\frac{H^{\frac{5}{2}}r^{2}}{f}\ddot{x}=0.
\end{equation}
Now we try to solve this problem for special case with the time
dependance $e^{-\mu t}$. In that case we put,
\begin{equation}\label{s38}
x(r, t)=x(r)e^{-\mu t},
\end{equation}
to equation (37) which yield to eigenvalue equation,
\begin{equation}\label{s39}
Ox=\mu^{2}x,
\end{equation}
where
\begin{equation}\label{s40}
O=\frac{f}{H^{\frac{5}{2}}r^{2}}\frac{d}{dr}\frac{fr^{2}}{\sqrt{H}}\frac{d}{dr}.
\end{equation}
Eigenvalue equation (39) at $\eta\rightarrow0$ limit  with $H=1$ and
by $\Lambda^{2}=-1$ reduces to associated Legendre equation which
have solutions in terms of hypergeometric functions. In other words
at such limit, our solutions are similar to solutions of the
following model, heavy quark moving through $\mathcal{N}$=4 super
Yang-Mills thermal plasma in 2 dimension [17]. As we expected
before, the calculation of drag force in the $\mathcal{N}$=2
supergravity theory at $\eta\rightarrow0$ limit is corresponding to
the heavy
quark in $\mathcal{N}$=4 super Yang-Mills theory.\\
In order to obtain $x(r)$ we expand it in terms of powers $\mu$ and
follow directly from [17], also take $\eta\rightarrow0$ to obtain,
\begin{eqnarray}\label{s41}
x_{1}^{\prime}(r)&=&\frac{-A}{(1+\Lambda^{2}r^{2})r^{2}},\nonumber\\
x_{2}^{\prime}(r)&=&\frac{A}{\Lambda^{2}}\left[\frac{1}{(1+\Lambda^{2}r^{2})r}-\frac{\pi}{2\Lambda(1+\Lambda^{2}r^{2})r^{2}}\right],
\end{eqnarray}
where $A$ ia an arbitrary constant. Then by using Neumann boundary
condition $x^{\prime}(r_{m})=0$ one can obtain friction coefficient
$\mu$ as a following,
\begin{equation}\label{s42}
\mu=\frac{\Lambda^{2}}{r_{m}-\frac{\pi}{2\Lambda}}.
\end{equation}
It is the lowest eigenvalue of operator $O$ or equivalently lowest
quasinormal modes of the string. The exponential form of $x$ at the
end point of string leads us to have $\dot{v}=-\mu v$, so, by using
the relation $m=T_{0}r_{m}$ at the $\eta\rightarrow0$ limit one can
obtain drag force as,
\begin{equation}\label{s43}
\frac{dp}{dt}=-\frac{T_{0}\Lambda^{2}r_{m}v}{r_{m}-\frac{\pi}{2\Lambda}}.
\end{equation}
Finally we obtain diffusion coefficient of the quark as,
\begin{equation}\label{s44}
D=\frac{T}{m\Lambda^{2}}(\frac{m}{T_{0}}-\frac{\pi}{2\Lambda}).
\end{equation}
Now by using equation (37) we will find total energy and momentum of
string. First, by using relation (12) and $-g\simeq\frac{1}{H}$ at
slow velocity we find,
\begin{equation}\label{s45}
\pi_{x}^{0}=-\frac{T_{0}}{\mu}(\frac{f
r^{2}}{\sqrt{H}}x^{\prime})^{\prime},
\end{equation}
in that case we have used from $x(r, t)=x(r)e^{-\mu t}$. Thus by
integrating from equation (45) and by Neumann boundary condition we
easily obtain  the total momentum as follows,
\begin{equation}\label{s46}
p=\frac{T_{0}}{\mu}\lambda(\Lambda,\beta,\eta)x^{\prime}(r_{min}),
\end{equation}
where,
\begin{equation}\label{s47}
\lambda(\Lambda,\beta,\eta)=\left[\frac{r_{min}^{2}+\Lambda^{2}r_{min}^{4}(1+\frac{\eta\sinh^{2}\beta}{r_{min}^{2}})^{3}-\eta}
{\sqrt{1+\frac{\eta\sinh^{2}\beta}{r_{min}^{2}}}}\right].
\end{equation}
and at $\eta\rightarrow0$ limit, $-g\rightarrow1$ we have,
\begin{equation}\label{s48}
p=\frac{T_{0}}{\mu}(1+\Lambda^{2}r_{min}^{2})r_{min}^{2}x^{\prime}(r_{min}),
\end{equation}
We see that at $\eta\rightarrow0$ limit the parameters $\beta$ and
$\eta$ play no role and we have
$\lambda(\Lambda)=1+\Lambda^{2}r_{min}^{2}$
in equation (48).\\
In order to obtain the total energy we expand $\frac{1}{\sqrt{-g}}$
to second order of speeds, then by using equation of motion we have,
\begin{equation}\label{s49}
\pi_{t}^{0}=-T_{0}\left[\frac{1}{\sqrt{H}}+\frac{1}{2}(\frac{f
r^{2}}{\sqrt{H}}xx^{\prime})^{\prime}\right].
\end{equation}
Here, by using Neumann boundary condition and integration on the
equation (49) we have,
\begin{equation}\label{s50}
E=T_{0}\left[\sqrt{r_{m}^{2}+\eta\sinh^{2}\beta}-\sqrt{r_{min}^{2}+\eta\sinh^{2}\beta}-
\frac{1}{2}\lambda(\Lambda,\beta,\eta)x(r_{min})x^{\prime}(r_{min})\right],
\end{equation}
which reduces to the following equation at the $\eta\rightarrow0$
limit,
\begin{equation}\label{s51}
E=T_{0}\left[r_{m}-r_{min}-
\frac{1}{2}(1+\Lambda^{2}r_{min}^{2})r_{min}^{2}x(r_{min})x^{\prime}(r_{min})\right].
\end{equation}
Let's take $p=M\dot{x}$ as momentum of particle with effective mass
$M$ and $x(r, t)=x(r)e^{-\mu t}$, so the equations (48) and (51)
give us the following equation,
\begin{equation}\label{s52}
E=T_{0}(r_{m}-r_{min})+\frac{p^{2}}{2M}.
\end{equation}
This is a well known relation between energy and momentum. By
comparing Equation (52) and (4) one can interpret
$T_{0}(r_{m}-r_{min})$ as $M_{rest}$ and $M$ as $M_{kin}$ of the
quark.
\section{Adding $B$-Field}
In this section, we consider a moving quark with constant velocity
$v$ through $\mathcal{N}$=2 supergravity thermal plasma, and
introduce a constant $B$-field including electric field $E$ and
magnetic field $\mathcal{H}$  along the $x^{1}$ and $x^{2}$
direction. Therefore one can couple $B$-field to line elements (8)
as following,
\begin{eqnarray}\label{s53}
ds^{2}&=&-\frac{f}{H^{2}}dt^{2}+H(r^{2}d\vec{x}^{2}+\frac{dr^{2}}{f})\nonumber\\
B&=&Edt\wedge dx_{1}+{\mathcal{H}}dx_{1}\wedge dx_{2},
\end{eqnarray}
where we have introduced the NS-NS constant antisymmetric fields $B_{01}=E$  and $B_{12}=\mathcal{H}$. All of other components of $B$-field are zero. Now we may describe such string attached to the quark by the following solutions,
\begin{eqnarray}\label{s54}
x_{1}(t, r)&=&x_{1}(r)+v_{1}t,\nonumber\\
x_{2}(t, r)&=&x_{2}(r)+v_{2}t,\nonumber\\
x_{3}(t, r)&=&0,
\end{eqnarray}
it means that string motion is in plan $(x_{1}, x_{2})$, so we have
velocity $\vec{\dot{x}}=(v_{1}, v_{2})$, and projected direction of
string tail ${\vec{x}}^{\prime}=(x_{1}^{\prime}, x_{2}^{\prime})$
which have different directions, these may be in opposite directions
or perpendicular to each other. As before, we choose static gauge
$\tau=t$ and $\sigma=r$, thus, the Lagrangian density is given by,
\begin{equation}\label{s55}
{\mathcal{L}}=-\sqrt{\frac{1}{H}-\frac{H^{2}
r^{2}}{f}{\vec{v}}^{2}+\frac{f
r^{2}}{H}{\vec{x}}^{\prime2}-\left(Ex_{1}^{\prime}+{\mathcal{H}}(\vec{v}\times{\vec{x}}^{\prime})\right)^{2}}.
\end{equation}
We continue our study in two cases of electric field and magnetic field separately.\\
First, we consider a moving quark with constant speed of $v$ and
constant electric field $E=B_{01}$. To study the effects of $E$, we
choose the moving direction of the quark in the $x^{1}$ direction.
Therefore we have a more restricted solution than the solutions (54)
as following,
\begin{eqnarray}\label{s56}
x_{1}(t, r)&=&x_{1}(r)+vt,\nonumber\\
x_{2}(t, r)&=&x_{3}(t, r)=0.
\end{eqnarray}
Now, by using Lagrangian density, we can calculate equation of
motion and $x_{1}(r)$ easily. Then $x_{1}$-component of momentum
density obtained by relation $\pi_{x_{1}}=\frac{\partial
{\mathcal{L}}}{\partial{x_{1}^{\prime}}}$. Finally by using the
reality and physical motion condition for $x_{1}(r)$ one can obtain,
\begin{equation}\label{s57}
\pi_{x_{1}}=\left(\frac{f(r_{c})r_{c}^{2}}{H(r_{c})}-E^{2}\right)^{\frac{1}{2}},
\end{equation}
where critical radius $r_{c}$ is the root of equation (25). So at
the $\eta\rightarrow0$ limit ($H=1$) where the critical radius
becomes $r_{c}^{2}=\frac{1}{v^{2}-\Lambda^{2}}$, one can obtain the
drag force as,
\begin{equation}\label{s58}
\frac{dp}{dt}=-T_{0}\left(\frac{v^{2}}{(v^{2}-\Lambda^{2})^{2}}+E^{2}\right)^{\frac{1}{2}},
\end{equation}
which is coincide with [27] if we set $\Lambda^{2}=1$ and
$1-v^{2}\equiv\frac{L^{4}}{r_{h}^{4}}$. In equation (58) we have
non-perturbation nature for the large $E$, however it may be
interesting to consider small $E$. In this case we expand square
root in equation (58) and find,
\begin{equation}\label{s59}
\frac{dp}{dt}=-T_{0}\frac{v}{(v^{2}-\Lambda^{2})}\left(1+\frac{(v^{2}-\Lambda^{2})^{2}}{2v^{2}}E^{2}+{\mathcal{O}}(E^{4})\right).
\end{equation}
Here, we note that the corrections are in terms of even power of
$E$. The effect of electric field on drag force is in the form of
$T_{0}\frac{\Lambda^{2}-v^{2}}{2v}E^{2}$ for small $E$. If $v>0$ and
$v^{2}>\Lambda^{2}$ or $v<0$ and $v^{2}<\Lambda^{2}$ then the
additional term due to electric field is opposite to drag force,
instead if $v<0$ and $v^{2}>\Lambda^{2}$ or $v>0$ and
$v^{2}<\Lambda^{2}$ then electric field increases the drag force.
Also if $E^{2}$ terms vanished,
we reproduce same results given in Refs. [17, 20].\\
Second, we consider the following solutions,
\begin{eqnarray}\label{s60}
x_{1}(t, r)&=&x_{1}(r)+vt,\nonumber\\
x_{2}(t, r)&=&x_{2}(r),\nonumber\\
x_{3}(t, r)&=&0.
\end{eqnarray}
Indeed we have ${\mathcal{H}}=B_{12}$ and $E=0$. Therefore, we have
two momentum density $\pi_{x_{1}}$ and $\pi_{x_{2}}$ conjugate with
coordinates $x_{1}$ and $x_{2}$ respectively. In this case one can
find,
\begin{eqnarray}\label{s61}
x_{1}^{\prime}&=&\pi_{x_{1}}\sqrt{\frac{\beta(\frac{1}{H}-\frac{H^{2}r^{2}}{f}v^{2})}{\frac{fr^{2}}{H}\left[(\pi_{x_{1}}^{2}-\frac{fr^{2}}{H})(\pi_{x_{2}}^{2}-\beta)-\pi_{x_{1}}^{2}\pi_{x_{2}}^{2}\right]}},\nonumber\\
x_{2}^{\prime}&=&\pi_{x_{2}}\sqrt{\frac{\frac{fr^{2}}{H}(\frac{1}{H}-\frac{H^{2}r^{2}}{f}v^{2})}{\beta\left[(\pi_{x_{1}}^{2}-\frac{fr^{2}}{H})(\pi_{x_{2}}^{2}-\beta)-\pi_{x_{1}}^{2}\pi_{x_{2}}^{2}\right]}},
\end{eqnarray}
where $\beta=\frac{fr^{2}}{H}-v^{2}{\mathcal{H}}^{2}$. The square
root quantity is positive for critical radius $r_{c}$ obtained by
equation (25), then, with respect to the solutions (60) we can find,
\begin{eqnarray}\label{s62}
\pi_{x_{1}}^{2}&=&(1+\frac{\eta \sinh^{2}\beta}{r_{c}^{2}})^{2}r_{c}^{4}v^{2},\nonumber\\
\pi_{x_{2}}^{2}&=&0,
\end{eqnarray}
which agrees with equation (34) where we obtained momentum density
without external field. Therefore one can obtain the drag forces as,
\begin{eqnarray}\label{s63}
\frac{dp_{1}}{dt}&=&-T_{0}(1+\frac{\eta \sinh^{2}\beta}{r_{c}^{2}})r_{c}^{2}v,\nonumber\\
\frac{dp_{2}}{dt}&=&0,
\end{eqnarray}
this means that there is no drag force along $x_{2}$-direction and the $B_{12}$  has no effect on the motion along $v_{1}$. Indeed $\frac{dp_{2}}{dt}=0$ agrees with a vanishing $v_{2}$ in solutions (60).\\
At the $\eta\rightarrow0$ limit, where $H=1$, we have
$r_{c}^{2}=\frac{1}{v^{2}-\Lambda^{2}}$ and
$\beta(r_{c})=v^{2}(\frac{1}{v^{2}-\Lambda^{2}}-{\mathcal{H}})$. In
this case one can find the drag forces as,
\begin{eqnarray}\label{s64}
\frac{dp_{1}}{dt}&=&-T_{0}\frac{v}{v^{2}-\Lambda^{2}},\nonumber\\
\frac{dp_{2}}{dt}&=&0.
\end{eqnarray}
Now we would like to consider the case of $\vec{v}\bot E$, where we
have the following solutions,
\begin{eqnarray}\label{s65}
x_{1}(t, r)&=&x_{1}(r),\nonumber\\
x_{2}(t, r)&=&x_{2}(r)+vt,\nonumber\\
x_{3}(t, r)&=&0,
\end{eqnarray}
instead solutions (60). We discuss at the $\eta\rightarrow0$ limit,
so, in this case we have the same analysis as above, but here there
are two solutions. The first solution gives $\pi_{x_{1}}=0$ and
$\pi_{x_{2}}=\sqrt{\beta(r_{c})}$. So we have,
\begin{eqnarray}\label{s66}
\frac{dp_{1}}{dt}&=&0,\nonumber\\
\frac{dp_{2}}{dt}&=&-T_{0}v\sqrt{\frac{1}{v^{2}-\Lambda^{2}}-{\mathcal{H}}}.
\end{eqnarray}
Therefore we have no drag force along the moving direction. Then,
the second solution gives $\pi_{x_{1}}=\frac{v}{v^{2}-\Lambda^{2}}$
and $\pi_{x_{2}}=0$. Hence we obtain the drag force same as in
equation (64). In this case the electric field $E$ has no effect on
the drag force.
\section{Higher Derivative Corrections}
The lowest order in the string length leads us to having an
expansion of the effective action in power of $\alpha^{\prime}$.
Higher powers of $\alpha^{\prime}$ correspond to higher derivative
terms provide information about naturally stringy effects, and
knowing them leads to some applications. The higher order
corrections are depend to the physics of  black hole. For instance,
they are relevant to the resolution of singularities of the black
hole. They are also significant for stretching the horizon if
solution of the classical black hole does not have one. They also
yield to a modification of the Beckenstein-Hawking area law for the
entropy [37]. Therefore higher derivative corrections allows us to
the better understand $AdS$/CFT correspondence. As we know in the
Ref. [34] the higher derivative Lagrangian is,
\begin{equation}\label{s67}
{\mathcal{L}}={\mathcal{L}}_{0}+{\mathcal{L}}_{R^{2}}+{\mathcal{L}}_{F^{4}}+{\mathcal{L}}_{RF^{2}},
\end{equation}
where ${\mathcal{L}}_{0}$ is given in equation (7) and the
additional terms are,
\begin{eqnarray}\label{s68}
e^{-1}{\mathcal{L}}_{R^{2}}&=&\alpha_{1}R^{2}+\alpha_{2}R_{\mu\nu}R^{\mu\nu}+\alpha_{3}R^{\mu\nu\rho\sigma}R_{\mu\nu\rho\sigma},\nonumber\\
e^{-1}{\mathcal{L}}_{F^{4}}&=&\beta_{1}(F_{\mu\nu}F^{\mu\nu})^{2}+\beta_{2}F^{\mu}_{\nu}F^{\nu}_{\rho}F^{\rho}_{\sigma}F^{\sigma}_{\mu},\nonumber\\
e^{-1}{\mathcal{L}}_{RF^{2}}&=&\gamma_{1}RF_{\mu\nu}F^{\mu\nu}+\gamma_{2}R_{\mu\nu}F^{\mu\rho}F_{\rho}^{\nu}+\gamma_{3}R^{\mu\nu\rho\sigma}F_{\mu\nu}F_{\rho\sigma}.
\end{eqnarray}
Here we consider the first order corrections to the $R$-charged
black hole solution of the equation (8) in Ref. [34]. They have
considered the coefficients  ($\alpha_{1}, \alpha_{2}, \cdots,
\gamma_{3}$) of the four-derivative terms in equation (68) as small
parameters. So, the first order corrections of charged black hole
solution (8) will be the following,
\begin{eqnarray}\label{s69}
A&=&\sqrt{3}\coth\beta(\frac{1+a_{1}}{H}-1)dt,\nonumber\\
f&=&1-\frac{\eta}{r^{2}}+\Lambda^{2}r^{2}H^{3}+f_{1}=f_{0}+f_{1},\nonumber\\
H&=&1+\frac{\eta \sinh^{2}\beta}{r^{2}}+h_{1}=h_{0}+h_{1},
\end{eqnarray}
where $a_{1}$, $f_{1}$ and $h_{1}$ are small corrections obtained in
Ref. [34]. We are going to apply the first order corrections of
charged black hole solution to drag force of quark, hence for the
moving quark through the thermal plasma with constant speed $v$ one
can write,
\begin{equation}\label{s70}
{\mathcal{L}}=-\sqrt{-g}=-\sqrt{\frac{1}{h_{0}+h_{1}}-\frac{(h_{0}+h_{1})^{2}
r^{2}}{(f_{0}+f_{1})}v^{2}+\frac{(f_{0}+f_{1})
r^{2}}{(h_{0}+h_{1})}{x^{\prime}}^{2}}.
\end{equation}
Similar to previous sections one can calculate drag force as,
\begin{equation}\label{s71}
\frac{dp}{dt}=-T_{0}r_{c}^{2}v(h_{0}+h_{1}).
\end{equation}
where $r_{c}$ is root of equation,
\begin{equation}\label{s72}
\frac{1}{h_{0}+h_{1}}-\frac{(h_{0}+h_{1})^{2}
r^{2}}{(f_{0}+f_{1})}v^{2}=0.
\end{equation}
To obtain expression (71) we assume that $r>>\eta$ but non-zero, so
one can neglect second powers of $\eta$ in correction terms
($a_{1}$, $h_{1}$ and $f_{1}$). In that case we have straightforward
generalization of relation (35) by replacing $H=h_{0}$ by
$H=h_{0}+h_{1}$. Here, the critical radius at the $\eta\rightarrow0$
limit reduces to the following equation,
\begin{equation}\label{s73}
r_{c}^{2}=\frac{1}{v^{2}-\Lambda_{eff}^{2}},
\end{equation}
where
\begin{equation}\label{s74}
\Lambda_{eff}^{2}=\Lambda^{2}\left[1+\frac{2}{3}(10\alpha_{1}+2\alpha_{2}+\alpha_{3})\Lambda^{2}\right],
\end{equation}
which $\alpha_{1}$, $\alpha_{2}$ and $\alpha_{3}$ are coefficients
related to the higher derivative corrections [34]. These are
specified by the underlying theory. So that the the $\alpha_{1}$ and
$\alpha_{2}$ coefficients are not fixed if we haven't an
off-mass-shell formulation such as string field theory, and they may
be vanished by an on-mass-shell field redefinition [34]. In this
way, we obtain physical information from the underlying string
theory just by the $\alpha_{3}$ coefficient. Thus, the drag force on
moving quark in the $\mathcal{N}$=2 supergravity thermal plasma at
the $\eta\rightarrow0$ limit and under higher derivative corrections
is,
\begin{equation}\label{s75}
\frac{dp}{dt}=-T_{0}\frac{v}{v^{2}-\Lambda_{eff}^{2}}.
\end{equation}
So the drag force coefficient $(\mu m)$ is given by
$\frac{T_{0}}{v^{2}-\Lambda_{eff}^{2}}$. Finally we can obtain
diffusion coefficient for the heavy quark as same as section 3, but
the cosmological constant  $\Lambda$ must replaced by the effective
cosmological constant $\Lambda_{eff}$. Similar to this, we can
generalize drag force in presence of electric and magnetic fields,
so direct extension of equations (58), (64) and (66) under higher
derivative corrections is given by replacing $\Lambda$ whit
$\Lambda_{eff}$.
\section{Conclusion}
By using $AdS$/CFT correspondence we studied the drag force on
moving quark through $\mathcal{N}$=2 supergravity thermal plasma. We
used the solutions of $AdS_{5}$ charged black hole and obtained
components of energy and momentum densities of string in three
cases; static string, moving straight string and moving curved
string. We have shown that the only physical state with constant
velocity is curved string. Also we discussed about quasinormal modes
of string and found drag force by lowest quasinormal modes of static
string which was accelerated. Then we considered $B$-field in the
thermal plasma and obtained the
effect of constant electric field and magnetic field on the drag force.\\
The interesting point and important result we obtained in this
article is that the limit of  $\eta\rightarrow0$ is corresponding to
the results of Ref. [17], where authors calculated energy loss of
heavy quark through $\mathcal{N}$=4 super Yang-Mills thermal plasma
by using $AdS$/CFT correspondence. Therefore, we summarized
corresponding parameter in table 1. So that at  $\eta\rightarrow0$
limit i.e. $H=1$ and with rescaling $r=L^{2}u$ and  setting $f\equiv
h$ for heavy quark both results of drag force in $\mathcal{N}$=2
supergravity theory and $\mathcal{N}$=4 super Yang-Mills theory are
similar. In another word at mentioned limit we obtained results of
Ref. [17] (see table 2).\\
Finally by considering higher derivative terms (first order
correction) in solutions (8) we obtained effect of such terms in
drag force on quark in $\mathcal{N}$=2 supergravity thermal plasma.
We found that at $\eta\rightarrow0$ limit, the drag force in various situations is generalized by replacing $\Lambda\rightarrow\Lambda_{eff}$.\\
For future works it is interesting to investigate $\eta\rightarrow0$
limit in other calculations such as jet quenching parameter [38-45]
or shear viscosity [46, 47, 48]. In that case one may find
correspondence between ${\mathcal{N}}$=2 supergravity and
${\mathcal{N}}$=4 super Yang-Mills theories.\\\\
\begin{center}
  \begin{tabular}{|@{} l @{} ||@{} c @{} | @{}r @{}|}
    \hline
    Quantity & $\mathcal{N}$=2 supergravity & $AdS$ \\ \hline\hline
    Minimum radius of D-brane & $\sqrt{(\frac{A_{+}}{T_{0}})^{2}+\frac{2A_{+}}{T_{0}}\sqrt{\frac{3Q}{2}}+\frac{Q}{2}}$ & $r_{m}$ \\ \hline
    Radius of horizon & $\sqrt{(\frac{B_{-}}{T_{0}})^{2}+\frac{2B_{-}}{T_{0}}\sqrt{\frac{3Q}{2}}+\frac{Q}{2}}$ & $r_{h}$ \\ \hline
    Lagrangian mass & $m$ & $T_{0}\left(r_{m}\sqrt{1+\frac{Q}{r_{m}^{2}}}-\sqrt{Q}\right)$ \\ \hline
    Thermal rest mass shift & $\Delta m(T)$ & $T_{0}\left(r_{h}\sqrt{1+\frac{Q}{r_{h}^{2}}}-\sqrt{Q}\right)$  \\ \hline
    Static thermal mass & $M_{rest}(T)$ & $T_{0}\left(r_{m}\sqrt{1+\frac{Q}{r_{m}^{2}}}-r_{h}\sqrt{1+\frac{Q}{r_{h}^{2}}}\right)$ \\
    \hline
  \end{tabular}
\end{center}
Table 1. $AdS$/CFT  translation table. Static thermal mass of quark,
$M_{rest}(T)$,  is equal to free energy of rest quark in
${\mathcal{N}}$=2 supergravity plasma which is Lagrangian mass of
quark $m$ at the limit of zero-temperature. We see that shift of
thermal rest mass $\Delta m(T)$ is depend on $r_{h}$. Here we define
$\eta\sinh^{2}\beta\equiv Q$, $M_{rest}(T)+\Delta m(T)\equiv A_{+}$ and $m-M_{rest}(T)\equiv B_{-}$.\\\\
\begin{center}
  \begin{tabular}{|l||c|r|}
    \hline
    Quantity ($\times L^{-2}$) & $\mathcal{N}$=4 SYM & $AdS$ \\ \hline\hline
    Minimum radius of D-brane  & $\frac{M_{rest}(T)+\Delta m(T)}{T_{0}}$ & $u_{m}$ \\ \hline
    Radius of horizon & $\frac{m-M_{rest}(T)}{T_{0}}$ & $u_{h}$ \\ \hline
    Lagrangian mass & $m$ & $T_{0}u_{m}$ \\ \hline
    Thermal rest mass shift & $\Delta m(T)$ & $T_{0}u_{h}$  \\ \hline
    Static thermal mass & $M_{rest}(T)$ & $T_{0}(u_{m}-u_{h})$ \\
    \hline
  \end{tabular}
\end{center}
Table 2.  $AdS$/CFT  translation table for heavy quark through
${\mathcal{N}}$=4 super Yang-Mills thermal plasma. Table 1. At the
$\eta\rightarrow0$ limit and rescaling $r=L^2 u$ reduces to the table 2.\\\\

{\bf Acknowledgments} It is pleasure to thank Anne Taormina and
Kasper Peeters for reading the manuscript and good suggestions for
K-K reduction.

\end{document}